\documentclass{article}
\usepackage{amsmath}
\usepackage{amsfonts}
\usepackage{amssymb}
\usepackage{braket}
\usepackage[utf8]{inputenc}
\usepackage[T1]{fontenc}
\usepackage[colorlinks=true]{hyperref}
\hypersetup{%
  colorlinks = true,
  linkcolor  = black
}
\usepackage{graphicx}
\usepackage[dvipsnames]{xcolor}
\usepackage{authblk}
\usepackage{soul}
\usepackage{cite}

\usepackage{amsthm}
\usepackage{comment}

\usepackage{ulem} 

\usepackage{enumitem}

\newtheorem{example}{Example}

\def\be{\begin{equation}}
\def\ee{\end{equation}}
\def\ba{\begin{eqnarray}}
\def\ea{\end{eqnarray}}

\def\Nl{{\mathchoice
{\setbox0=\hbox{$\displaystyle\rm N$}\hbox{\hbox to0pt
{\kern0.4\wd0\vrule height0.9\ht0\hss}\box0}}
{\setbox0=\hbox{$\textstyle\rm N$}\hbox{\hbox to0pt
{\kern0.4\wd0\vrule height0.9\ht0\hss}\box0}}
{\setbox0=\hbox{$\scriptstyle\rm N$}\hbox{\hbox to0pt
{\kern0.4\wd0\vrule height0.9\ht0\hss}\box0}}
{\setbox0=\hbox{$\scriptscriptstyle\rm N$}\hbox{\hbox to0pt
{\kern0.4\wd0\vrule height0.9\ht0\hss}\box0}}}}
\def\Zl{{\mathchoice
{\setbox0=\hbox{$\displaystyle\rm Z$}\hbox{\hbox to0pt
{\kern0.4\wd0\vrule height0.9\ht0\hss}\box0}}
{\setbox0=\hbox{$\textstyle\rm Z$}\hbox{\hbox to0pt
{\kern0.4\wd0\vrule height0.9\ht0\hss}\box0}}
{\setbox0=\hbox{$\scriptstyle\rm Z$}\hbox{\hbox to0pt
{\kern0.4\wd0\vrule height0.9\ht0\hss}\box0}}
{\setbox0=\hbox{$\scriptscriptstyle\rm Z$}\hbox{\hbox to0pt
{\kern0.4\wd0\vrule height0.9\ht0\hss}\box0}}}}
\def\Ql{{\mathchoice
{\setbox0=\hbox{$\displaystyle\rm Q$}\hbox{\hbox to0pt
{\kern0.4\wd0\vrule height0.9\ht0\hss}\box0}}
{\setbox0=\hbox{$\textstyle\rm Q$}\hbox{\hbox to0pt
{\kern0.4\wd0\vrule height0.9\ht0\hss}\box0}}
{\setbox0=\hbox{$\scriptstyle\rm Q$}\hbox{\hbox to0pt
{\kern0.4\wd0\vrule height0.9\ht0\hss}\box0}}
{\setbox0=\hbox{$\scriptscriptstyle\rm Q$}\hbox{\hbox to0pt
{\kern0.4\wd0\vrule height0.9\ht0\hss}\box0}}}}
\def\Rl{{\mathchoice
{\setbox0=\hbox{$\displaystyle\rm R$}\hbox{\hbox to0pt
{\kern0.4\wd0\vrule height0.9\ht0\hss}\box0}}
{\setbox0=\hbox{$\textstyle\rm R$}\hbox{\hbox to0pt
{\kern0.4\wd0\vrule height0.9\ht0\hss}\box0}}
{\setbox0=\hbox{$\scriptstyle\rm R$}\hbox{\hbox to0pt
{\kern0.4\wd0\vrule height0.9\ht0\hss}\box0}}
{\setbox0=\hbox{$\scriptscriptstyle\rm R$}\hbox{\hbox to0pt
{\kern0.4\wd0\vrule height0.9\ht0\hss}\box0}}}}
\def\Cl{{\mathchoice
{\setbox0=\hbox{$\displaystyle\rm C$}\hbox{\hbox to0pt
{\kern0.4\wd0\vrule height0.9\ht0\hss}\box0}}
{\setbox0=\hbox{$\textstyle\rm C$}\hbox{\hbox to0pt
{\kern0.4\wd0\vrule height0.9\ht0\hss}\box0}}
{\setbox0=\hbox{$\scriptstyle\rm C$}\hbox{\hbox to0pt
{\kern0.4\wd0\vrule height0.9\ht0\hss}\box0}}
{\setbox0=\hbox{$\scriptscriptstyle\rm C$}\hbox{\hbox to0pt
{\kern0.4\wd0\vrule height0.9\ht0\hss}\box0}}}}
\def\Hl{{\mathchoice
{\setbox0=\hbox{$\displaystyle\rm H$}\hbox{\hbox to0pt
{\kern0.4\wd0\vrule height0.9\ht0\hss}\box0}}
{\setbox0=\hbox{$\textstyle\rm H$}\hbox{\hbox to0pt
{\kern0.4\wd0\vrule height0.9\ht0\hss}\box0}}
{\setbox0=\hbox{$\scriptstyle\rm H$}\hbox{\hbox to0pt
{\kern0.4\wd0\vrule height0.9\ht0\hss}\box0}}
{\setbox0=\hbox{$\scriptscriptstyle\rm H$}\hbox{\hbox to0pt
{\kern0.4\wd0\vrule height0.9\ht0\hss}\box0}}}}
\def\Ol{{\mathchoice
{\setbox0=\hbox{$\displaystyle\rm O$}\hbox{\hbox to0pt
{\kern0.4\wd0\vrule height0.9\ht0\hss}\box0}}
{\setbox0=\hbox{$\textstyle\rm O$}\hbox{\hbox to0pt
{\kern0.4\wd0\vrule height0.9\ht0\hss}\box0}}
{\setbox0=\hbox{$\scriptstyle\rm O$}\hbox{\hbox to0pt
{\kern0.4\wd0\vrule height0.9\ht0\hss}\box0}}
{\setbox0=\hbox{$\scriptscriptstyle\rm O$}\hbox{\hbox to0pt
{\kern0.4\wd0\vrule height0.9\ht0\hss}\box0}}}}
\newcommand{\ct}{\mathcal T}

\newcommand{\eqa}{\begin{eqnarray}}
\newcommand{\neqa}{\end{eqnarray}}

\definecolor{myblue}{rgb}{0.2,0.2,0.8}

\usepackage{bm}
\usepackage{bbm}

\newcommand{\lalg}[1]{\mathfrak{#1}}  

\newcommand{\su}{\lalg{su}}
\renewcommand{\sl}{\lalg{sl}}

\newcommand{\ketbra}[2]{|#1 \rangle \!\langle #2 |}

\def\I{{\mathbb I}}
\def\A{{\sf A}}
\def\B{{\sf B}}

\def\sl2c{{\mathfrak{sl}_2 \Cl}} 
\def\su2{{\mathfrak{su}_2}}

\definecolor{darkgreen}{rgb}{0.0, 0.5, 0.13}

\usepackage{geometry}
\begin{document}

\title{Reply to ``Masanes-Galley-M\"uller and the State-Update Postulate''}

\author[1]{Thomas D. Galley\thanks{\texttt{thomas.galley@oeaw.ac.at}}}
\author[2,3]{Llu\'is Masanes}
\author[1,4,5]{Markus P.\ M\"uller}
\affil[1]{\small Institute for Quantum Optics and Quantum Information, Austrian Academy of Sciences,\newline Boltzmanngasse 3, 1090 Vienna, Austria}
\affil[2]{\small Department of Computer Science, University College London, United Kingdom}
\affil[3]{\small London Centre for Nanotechnology, University College London, United Kingdom}
\affil[4]{\small Perimeter Institute for Theoretical Physics, 31 Caroline Street North, Waterloo ON N2L 2Y5, Canada}
\affil[5]{\small Vienna Center for Quantum Science and Technology (VCQ), Faculty of Physics, University of Vienna, Boltzmanngasse 5, 1090 Vienna, Austria}

\date{December 7, 2022}

\maketitle

\begin{abstract}
In a recent comment on the arXiv, Blake C.\ Stacey criticizes our derivation of the quantum state update rule in Nat.\ Commun.\ \textbf{10}, 1361 (2019). Here we argue that the criticism is unfounded. In particular, and in contrast to Stacey's claims, our proof does not assume linearity.
\end{abstract}

\section{Introduction}

In a recent comment on the arXiv~\cite{stacey_2022}, Blake Stacey offers a critique of our work~\cite{Masanes_2019}. The results of~\cite{Masanes_2019} demonstrate that the measurement postulates of quantum theory are the only  possible ones consistent with the structure of pure states, unitary dynamics and the composition rule of quantum theory. Stacey's critique is not levelled at the main theorem (the Born rule), but rather at its corollary (the post-measurement state-update rule.) Stacey claims that our proof of the state-update rule is circular, allegedly assuming the linearity that it intends to show.

In the following, we provide an overview of our paper~\cite{Masanes_2019} and outline Stacey's argument from~\cite{stacey_2022} before offering our response. 

\section{Overview of~\cite{Masanes_2019}}

The quantum measurement postulates state that each measurement is mathematically characterised by a POVM, and that the probability of measurement outcomes is given by the Born rule. The main theorem of~\cite{Masanes_2019} shows that the quantum measurement postulates are the only operationally consistent postulates given that one assumes the `rest of' quantum theory; namely that pure states are represented by rays in some complex projective space, reversible dynamics by unitary transformations, and pure states compose using the tensor product. 

We stress that for single systems (i.e. without any constraints on the composition of systems) this is not the case: there is an infinite family of conceivable systems with the same pure states and reversible dynamics as quantum theory but different measurement rules~\cite{Galley_2017}. A key point to emphasise is that these alternative systems have sets of mixed states which do not correspond to the set of density operators, and sets of outcome probability functions which do not correspond to the set of POVM elements.

The density operator already contains information about the measurements of quantum theory since density operators tell us which ensembles of states are indistinguishable. Here indistinguishability is relative to the possible measurements. Mathematically, this is evident by the fact that if one assumes that the states of a system are given by density operators (with statistical mixing represented by convex combinations), it is immediate that the measurements must be represented by POVMs and probabilities by the Born rule. As Stacey points out, this type of argument goes back at least to Holevo.

Thus, in order to be non-trivial, any proof of the Born rule must not assume that probabilities are linear in density operators (or in POVM elements for similar reasons).
The starting point of the theorem of~\cite{Masanes_2019} is that a measurement outcome should be some function of the pure states, namely of the form $\mathbf{f}(\psi)$. This function need not be linear in $\ketbra{\psi}{\psi}$ (in fact, it was proven in \cite{Galley_2017} that $F$ has to be linear in $\ketbra{\psi}{\psi}^{\otimes n}$, for some fixed integer $n\geq 1$.)
Thus \textit{the main theorem of~\cite{Masanes_2019} neither assumes that the outcome probability functions are linear in $\ketbra{\psi}{\psi}$ nor that outcomes correspond to POVM elements} but rather \textit{derives} these facts. 

In~\cite{Masanes_2019} it is also shown that, once the measurement postulates are known to be the quantum ones, the only consistent state update rule is the quantum one. The argument for this follows from standard arguments and can be found in the proof of Lemma 22 in the supplementary materials of~\cite{Masanes_2019}.

\section{Stacey's argument}

The proof that the measurement update rule is given by the standard form requires introducing a POVM $\{H_{j,i}\}_{j,i}$ which corresponds to the sequence of two POVMs $\{F_i\}_i$ and $\{G_j\}_j$. Stacey argues that, although the main theorem of~\cite{Masanes_2019} applies to single-step measurements, it does not apply to two step transformations. He states:

\begin{quote}
    In other words, the linearity of the map $
    \Lambda_{F_i}(\rho)$ follows from the assumption that $H_{i,j}$ is a POVM element that can be used as any other. But $\{H_{j,i}\}$ is not a measurement that is performed at a single time. It is, by necessity, a sequence of two experimental interventions, with a state-change in between.
\end{quote}

His main point of contention is that, while it may be justified in assigning POVM elements to the individual measurements outcomes of $\{F_i\}_i$ and $\{G_j\}_j$, it is not automatically justified to assert that the outcomes $\{(F_i, G_j)\}_{j,i}$ corresponding to the sequential measurement can be associated to a single POVM $\{H_{j,i}\}_{j,i}$.
That is, in order to associate a POVM elements to the outcomes of a sequence of measurements we must have made some assumptions about sequential measurements:

\begin{quote}
     the assumption that two-stage
experiments have their mathematical representations in exactly the same set as one-stage experiments is an assumption that state-change is undramatic. (We have already accustomed
ourselves to the fact that the AND operation is not always meaningful for quantum experiments; on what deep basis do we say it should apply trivially here?)
\end{quote}

To back up his argument he provides the following example where a sequence of measurements  $\{F_i\}$ and $\{G_j\}$ would not lead to a POVM elements $\{H_{i,j}\}$ being associated to the sequence:

\begin{quote}
    We could have a first measurement  $\{F_i\}$ and a second measurement  $\{G_j\}$, but the transformation taking place in between would be nonlinear, and so the conclusion about the map $\Lambda_{F_i}$ would simply not follow. For instance, suppose the initial state $\rho$ is a mixture of the Pauli eigenstate $\ketbra{+z}{+z}$ and the garbage state $\frac{\I}{2}$, and let the state-update map implement a logistic transformation of the Bloch-sphere radial coordinate: $r \to r'= \lambda r (1-r)$ with $\lambda \in [0,4]$. If the first and second POVMs are both the computational basis measurement
$\{\ketbra{+z}{+z}, \ketbra{-z}{-z\}}$, then the probability of either outcome in the second measurement is trivially a nonlinear function of the initial state $\rho$.
\end{quote}

The measurement with outcomes $\{(F_i,G_j)\}_{i,j}$ would in this case not be a linear function of $\rho$ and would therefore not  correspond to a POVM $\{H_{j,i}\}_{j,i}$ with probabilities given by the Born rule.
Therefore, Stacey concludes that we have therefore implicitly assumed linearity of the state update rule in the proof of the linearity of the state update rule.

\section{Response}

The key point of our response is that \textit{any experiment that takes a quantum system, described by a Hilbert space $\mathbb{C}^d$, as its input and generates one of several possible outcomes} is a measurement. Our derivation of the Born rule is completely agnostic as to how the outcome comes about or how many steps it takes.

More specifically, our result is a mathematical theorem about the form of what we call ``outcome probability functions'' (OPFs) $\mathbf f:\mathrm{P}\mathbb{C}^d\to[0,1]$, $\psi\mapsto \mathbf  f(\psi)$. Every $\mathbf f$ that satisfies a list of assumed mathematical properties (given in our paper) is then proven to be of the form $\psi\mapsto\langle\psi|Q|\psi\rangle$, where $Q$ is some POVM element. \text{Physically,} our result will thus apply to all experiments in which the mathematical properties are enforced by the scenario at hand. Such experiments can be thought of as consisting of two stages:
\begin{itemize}
    \item The \textbf{preparation} P of a quantum state $\psi$. Note that our result is agnostic as to what the state means --- it could be understood as the real state of affairs of the physical system that exits a well-characterized preparation device; or an agent's formulation of their beliefs about future experiences involving the quantum system (as QBists might prefer), or something completely different.
\item The \textbf{rest} R of the experiment, generating an \textbf{outcome} to which we want to assign probabilities.
\end{itemize}
Our assumed mathematical properties correspond to physical assumptions on how our experiment relates to other experiments that we could possibly perform instead (or in addition), given the experiment that we have. For example, Property 2 of~\cite{Masanes_2019} assumes that if $\mathbf f$ is an OPF, and if $U$ is a unitary, then $\mathbf f\circ U$ is also a valid OPF. Operationally, this means the following: after performing preparation P, we could also (if we wanted) implement  transformation $U$ on the system, before we complete the experiment by performing R. If we know how to assign probabilities to the possible outcomes of our original experiment (via $\mathbf f$), then we also know how to assign probabilities to the possible outcomes of this new experiment (namely via $\mathbf f\circ U$). Hence $\mathbf f\circ U$ must be a valid OPF (note that this argumentation does \textit{not} involve any assumptions about the \textit{mathematical structure} of $\mathbf f$, like being linear on $|\psi\rangle \!\langle\psi|$, etc.).

Similarly, our Property 1 (closedness under statistical mixing) formalizes the physical possibility that if we have a finite set of experiments (labelled by $x$) with identical preparation procedures, but different outcome-generating stages $R_x$, then we obtain a new valid experiment by performing experiment $x$ with some probability $p_x$.

Note that \textit{a very large class of experiments} satisfies the above desiderata, i.e.\ is in the realm of applicability of our derivation of the Born rule.
\begin{example}
Consider an experiment where the preparation $P$ amounts to generating an entangled state $\psi$ of the spin degrees of freedom of two electrons. The outcome-generating stage $R$ of the experiment works as follows. We measure whether electron $1$ has spin-up in our favorite $z$-direction. If so, then the outcome of our experiment is by definition the string of five zeroes $00000$. Otherwise, we repeatedly measure electron $2$ in the x-,z-,x-,z-,x-basis, denoting `0' for `up' and `1' for `down'. If the resulting string $s$ contains less than four zeroes, then this is the outcome string. Otherwise, we wait until the next day and repeat the procedure. And so on, ad infinitum, until we terminate.

This experiment falls into the regime of applicability of our derivation of the Born rule. The probabilities of the outcomes must therefore be given by some POVM.
\end{example}
However, there are certainly situations to which our mathematical theorem does not apply:
\begin{example}
Imagine an experiment on a quantum system $S$ which is constrained by a fundamental superselection rule, forbidding superpositions between some charge sectors. In this case, for some unitaries $U$, it will be fundamentally impossible to implement them on $S$. This eradicates the motivation for Property 2 as explained above: the physical scenario does not motivate the mathematical assumption that $\mathbf{f}\circ U$ is a valid OPF for every unitary $U$ and every OPF $\mathbf{f}$. Alternatively, we might say that the description of $S$ as a Hilbert space $\mathbb{C}^d$ has been incomplete in the first place. Thus, our theorem is inapplicable and cannot be used to derive the Born rule for $S$.

Similarly, we may think of Wigner's-friend-type situations with a sequence of outcomes that, intuitively, do not necessarily ``co-exist''~\cite{Bong_2020} so that it is unclear whether, and how, it makes sense to assign (joint) probabilities to these sequences of outcomes. The question of applicability of our theorem may then depend on details of the metaphysical interpretation of such scenarios.
\end{example}

The main theorem of~\cite{Masanes_2019} shows that outcomes of measurements that satisfy the above desiderata must correspond to POVM elements with probabilities given by the Born rule (which is linear in $|\psi\rangle \!\langle \psi|$). The theorem does not assume linearity at the level of density operators or POVM elements, but instead, it derives it.

In particular, a sequence of measurements  $\{F_i\}_i$ and $\{G_j\}_j$  can be interpreted as an outcome-generating procedure which inputs a state $\rho$ and outputs an outcome pair $(i,j)$. The fact that this outcome set has a direct product structure or that the parts of the outcome pair are obtained sequentially (and potentially in some exotic manner) has no bearing on the fact that an outcome $(i,j)$ is still the outcome of a measurement that falls into the regime of physical applicability of our mathematical theorem, in every case in which it makes sense to talk about the probability $p(i,j)$ of the pair of outcomes $(i,j)$. Similarly two measurements 
 $\{F_i^\A\}_i$ and $\{G_j^\B\}_j$  performed on subsystems $\A$ and $\B$ respectively, also form a single measurement of the composite system $\A\B$.

Let us return to Stacey's example and show how it can be ruled out by appealing to our operational characterisation of measurements as any procedure which inputs a system and outputs some probability distribution over classical variables.

It is clear from this that if a non-linear transformation $\ct$ were to occur in between step 1 and step 2 of a sequential measurement as above, the operation $\{G_j \circ \ct\}_j$ would correspond to a measurement operationally. Indeed it is a single step measurement (in the language of Stacey), and since the outcome probabilities are non-linear in $\ketbra{\psi}{\psi}$, it is inconsistent with the other postulates of quantum theory by the main theorem.

\begin{quote}
    The existence of the $H_{i,j}$ on which MGM’s lemma depends follows from the linearity they
wish to prove. This linearity can be derived from other premises, such as assumptions about
context-independence [4] as in the proofs of the POVM-Gleason theorem [5, 6]. However,
these assumptions do not appear to be entailed in the postulates that MGM put forth.
Indeed, MGM write, “We stress that our results, unlike previous contributions [ . . . ], do not
assume” such premises.
\end{quote}

We stress   that the linearity of $H_{i,j}$ follows from the main theorem of~\cite{Masanes_2019} which does not make any assumption of linearity.  

We also want to mention that, independently of its tension with the main theorem of~\cite{Masanes_2019}, the consistency of Stacey's example is not clear to us. The example is formulated in terms of density matrices, as if these fully characterise a state, or the outcome probabilities when a state is measured. However, the non-linear dynamics of his example-theory would allow for discriminating two ensembles of the same mixed state. This implies that the density matrix does not provide a full characterisation the outcome probabilities when a state is measured.

\bibliographystyle{utphys}

\end{document}